\documentclass[aps,preprint,nofootinbib]{revtex4}
\usepackage{amsmath}
\usepackage{graphicx}
\usepackage{color}
\usepackage{ulem}
\usepackage{slashed}
\usepackage{subfigure}
\usepackage{tabularx}
\usepackage{booktabs}
\usepackage{mathtools}
\usepackage{hyperref}
\usepackage{multirow}
\numberwithin{figure}{section}
\allowdisplaybreaks[4]

\newcommand{\dlog}{{\text{Li}_2}}

\begin{document}

\title{NLO Corrections to Dimuonium Production in Photon-Photon Collision\\[0.7cm]}

\author{\vspace{1cm} Qi-Ming Feng$^{1}$, Si-Mao Guo$^{1,2}$, Qi-Wei Hu$^{1}$, Cong-Feng Qiao$^{1,3}$\footnote[1]{qiaocf@ucas.ac.cn}, Qi-Ming Qiu$^{3}$, Yu-Jie Tian$^{1,2}$, Bai-Rong Zhang$^{3}$, Hao Zhang$^{2,1,4}$\footnote[1]{zhanghao@ihep.ac.cn}, Xuan-Heng Zhang$^{1}$, and Bo-Ting Zhou$^{1,2}$}

\affiliation{\small {$^1$ School of Physics, University of Chinese Academy of
Sciences, Beijing 100049, China\\
$^2$ Institute of High Energy Physics, Chinese Academy of Sciences, Yuquan Road 19B, Beijing 100049, China\\
$^3$ International Centre for Theoretical Physics Asia-Pacific, University of Chinese Academy of Sciences, Beijing 100190, China\\
$^4$ Center for High Energy Physics, Peking University, Beijing 100871, China}
}

\author{~\\}

\begin{abstract}
\vspace{0.5cm}
Dimuonium ($\mu^+ \mu^-$) is one of the pure QED bound systems of leptons, together with positronium and ditauonium. The former had been observed in experiment in 1951, while the search for dimuonium and ditauonium  ($\tau^+ \tau^-$) are still in vain. The ditauonium is thought to be hard to measure due to its short life time, whereas the dimuonium is very likely to be observed in current running experiments. We calculate in this work the para and ortho dimuonium production in photon-photon collision at the next-to-leading order (NLO) in QED. To handle the Coulomb divergence within the bound system, the non-relativistic QED description is adopted for consistency. The results indicate that the NLO corrections are negative for both para and ortho states, at the normal order of perturbative QED corrections. The measurement of the dimuonium in Belle II experiment, especially at the forthcoming facility of STCF, is tenable.

\end{abstract}
\maketitle

\newpage

\section{Introduction}

In analogous to atom, a pure lepton pair may form a bound state, the leptonium, owing to the electromagnetic interaction. The lightest leptonium, the $e^+ e^-$ positronium, was predicted \cite{Pirenne1946} and observed many decades ago \cite{Deutsch1951}, while the other two pure leptonium triplets $\mu^+ \mu^-$ and  $\tau^+ \tau^-$ are still missing. Since the $\tau^\pm$ are relatively heavy and have a large decay width via weak interaction, the $\tau^+ \tau^-$ bound system is short-lived, which makes it hard to be observed \cite{Brodsky2009}. Considering the $\mu^+ e^-$ bound state has been coined the muonium earlier, $\mu^+ \mu^-$ and  $\tau^+ \tau^-$ systems are usually called pure muonium and tauonium, or dimuonium and ditauonium, respectively. In this work, for simplicity, we will adopt the latter nomenclature. After the discovery of muonium ($\mu^+ e^-$) in experiment in 1960 \cite{Hughes1960}, the next compact pure electromagnetic bound system to be observed should be the dimuonium ($\mu^+ \mu^-$), though obviously tedious, that has been theoretically investigated in depth for a long time \cite{Bilenkii1969,Hughes1971,Malenfant1987,Karshenboim1998} (cf. \cite{Mohsen2015} for a more comprehensive review). Hereafter, the dimuonium will be denoted by $D_m$ for short in our discussion.

Though one can in principle evaluate the production and decays of leptonic bound states by virtue of QED, as most of works did before, to systematically truncate the calculation in orders of relativistic and electromagnetic coupling expansions, the theory of Nonrelativistic QED (NRQED) \cite{Caswell1986} behaves as an even physical enlightening and calculation effective theoretical framework to this aim. In practice, in dealing with the electromagnetic bound states, NRQED framework is even simpler than utilizing the Bethe-Salpeter equation. In this paper, we will employ the NRQED framework to categorize different contributions in the calculation of dimuonium production at leading order (LO) relativistic expansion in relative velocity of leptons and the next-to-leading order (NLO) in QED. 

The ground states of a di-muon system are in spin singlet and triplet, viz para state $D_m(^1S_0)$ and ortho state $D_m(^3 S_1)$, respectively. For $D_m(^1S_0)$ the dominant decay mode is to two photons, while $D_m(^3S_1)$ predominantly decays to two electron-positron pairs. Theoretical calculations tell that $D_m(^1S_0)$ has a lifetime of 0.602 ps \cite{Bilenkii1969,Hughes1971} and $D_m(^3 S_1)$'s lifetime is about 1.81 ps \cite{Bilenkii1969}, which are relatively long in the regime of metastable atoms. Recently, Gargiulo, Meco and Palmisano give a detailed LO analysis on the feasibility of observing $D_m(^1S_0)$ via photon-photon interaction with collected Belle II data \cite{Gargiulo2025}, and find that the signal significance may exceed 5$\sigma$ confidence level (C.L.) over the dominant background. In this work, we calculate the NLO QED corrections to $D_m(^1S_0)$ and $D_m(^3 S_1)$ production in photon-photon collision. The paper is organized as follows. In section II, we present the formalism employed and some formulations. Section III exhibits some numerical results, illustrating the observation potential of dimuonium production in photon-photon collision. In the last section, we make a conclusion of our calculation and give some comments on the results.

\section{Formalism and Calculation}

\begin{figure}[htb]
    \includegraphics[width=13 cm]{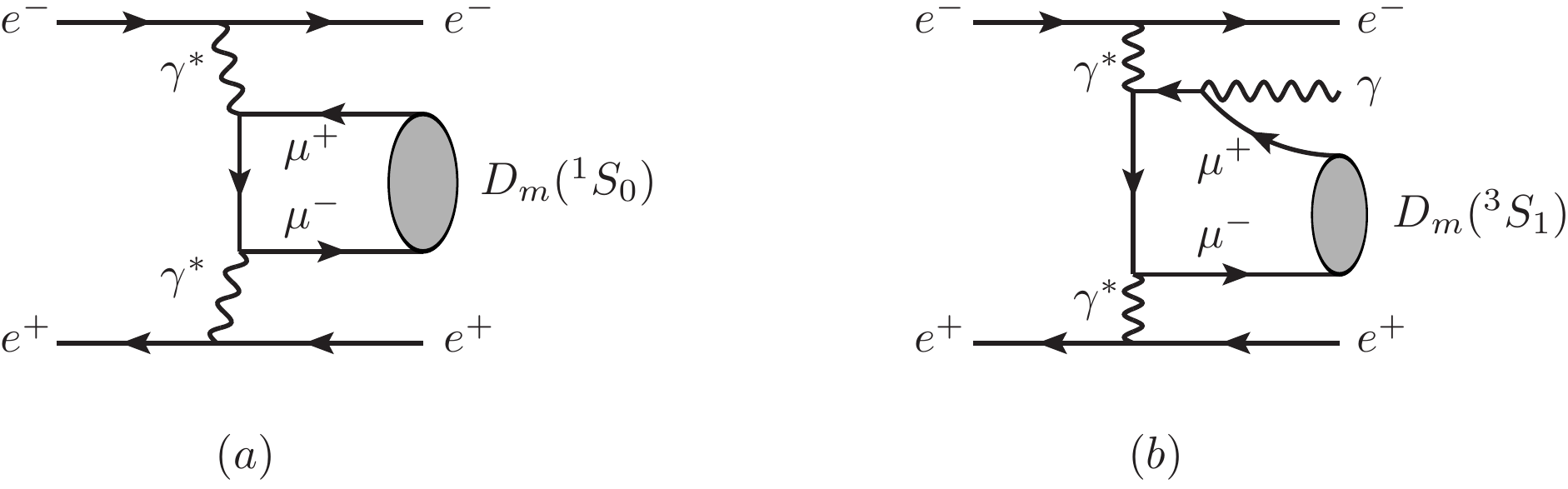}
    \caption{Generic LO Feynman diagrams of para (a) and ortho (b) dimuonium photoproduction in electron-positron collision. }\label{fig_Feynlo}
\end{figure}

In high energy experiment, the photon-photon scattering can be achieved at $e^+ e^-$ collider like SuperKEKB, where the initial photons are generated via bremsstrahlung effect, as shown in Figure \ref{fig_Feynlo}. By taking the equivalent photon approximation (EPA) \cite{Budnev1975}, the intermediate photons can be treated as real. The energy spectrum of bremsstrahlung photon is well formulated in Weizsacker-Williams approximation (WWA) \cite{Frixione:1993yw}:
\begin{equation}
    f_{\gamma}(x) = \frac{\alpha}{2\pi}\left[\frac{1+(1-x)^2}{x}\log\left(\frac{Q^{2}_{\rm max}}{Q^{2}_{\rm min}}\right)+2m^2_{e}x\left(\frac{1}{Q^{2}_{\rm max}}-\frac{1}{Q^{2}_{\rm min}}\right)\right],
\end{equation}
where $Q^{2}_{\rm min}=m^{2}_{e}x^{2}/(1-x)$, $Q^{2}_{\rm max}=(\theta_{c}\sqrt{S}/2)^2(1-x)+Q^{2}_{\rm min}$, $m_e$ is the electron mass, $x=E_{\gamma}/E_{e}$ is the energy fraction of the photon, $\sqrt{S}$ is the collision energy of the $e^+e^-$ collider, $\theta_{c}=32$ mrad is the maximum scattering angle of the electron or positron \cite{Klasen:2001cu}.

The total cross section of dimuonium production in photon-photon collision can then be obtained by convoluting the $\gamma+\gamma\to {Q_m}[\mu^+ \mu^-] + X $ cross section with the photon distribution functions:
\begin{equation}
    d\sigma=\int dx_{1}dx_{2} f_{\gamma}(x_{1})f_{\gamma}(x_{2})d \hat{\sigma}( \gamma +\gamma\to {Q_m}[\mu^+ \mu^-] + X) \ ,
\end{equation}
where ${Q_m}$ represents para and othro dimuonium, and $X$ denotes the possible final state photon.
In this work, the photon-photon scattering cross sections $d \hat{\sigma}$ will be calculated perturbatively up to the NLO in QED. It can be schematically expressed as
\begin{equation}
    d\hat{\sigma}( \gamma+\gamma\to {Q_m}[\mu^+ \mu^-] + X) = d\hat{\sigma}_{\rm born} + d\hat{\sigma}_{\rm virtual} + d\hat{\sigma}_{\rm real} + \mathcal{O}(\alpha^{4})\ .
\end{equation}
The Born level cross section, the virtual correction, and the real correction respectively take the following forms:
\begin{equation}
    \begin{split}
        &d\hat{\sigma}_{\rm born}=\frac{1}{2 {s}}\overline{\sum}|\mathcal{M}_{\rm tree}|^{2}d{\Phi}_{2}\ ,\\
        &d\hat{\sigma}_{\rm virtual}=\frac{1}{2 {s}}\overline{\sum}2{\rm Re}(\mathcal{M}^{*}_{\rm tree}\mathcal{M}_{\rm oneloop})d{\Phi}_{2}\ ,\\
        &d\hat{\sigma}_{\rm real}=\frac{1}{2 {s}}\overline{\sum}|\mathcal{M}_{\rm real}|^{2}d{\Phi}_{3}\ ,
    \end{split}
\end{equation}
where the small letter ${s}$ represents the center-of-mass (c.m.) energy squared of two photons, $\overline{\sum}$ means sum (average) over the polarizations of initial and final state particles, $d{\Phi}_{2}$ ($d{\Phi}_{3}$) denotes final state two (three)-body phase space.

The LO cross sections of  $D_m(^1S_0)$ and $D_m(^3 S_1)$ production in photon-photon collision are pretty concise, they are
\begin{equation}
d\hat{\sigma} = \frac{64 \pi^2 \alpha^2}{m}|{\Psi}(0)|^2 d \Phi_1\ , 
\label{1S0CS}
\end{equation}
and
\begin{equation}
d\hat{\sigma}=\frac{2048 \pi^2 \alpha^3 m\left[(s^2+t^2)(s+t-4m^2)^2+st(s-4m^2)(t-4m^2)\right] }{s^2 (s - 4 m^2)^2 (t - 4 m^2)^2 (s + t)^2}|{\Psi}(0)|^2d\Phi_2 \ ,
\end{equation}
respectively. Here, $t$ is the standard Mandelstam variable in two-to-two scattering; $|{\Psi}(0)|^2$ is the squared wavefunction at the origin of dimuonium. At the LO in relativistic expansion, both para and ortho states have the same value in ${\Psi}(0)$, which can be determined by solving the non-relativistic Schr\"odinger equation with Coulomb potential. It writes
\begin{equation}
|\Psi(0)|^2 = \frac{(\alpha m_\mu)^3}{8 \pi} ,
\end{equation}
with $\alpha$ the fine structure constant and $m_\mu$ the muon mass. 

\begin{figure}[htb]
    \includegraphics[width=0.85\textwidth]{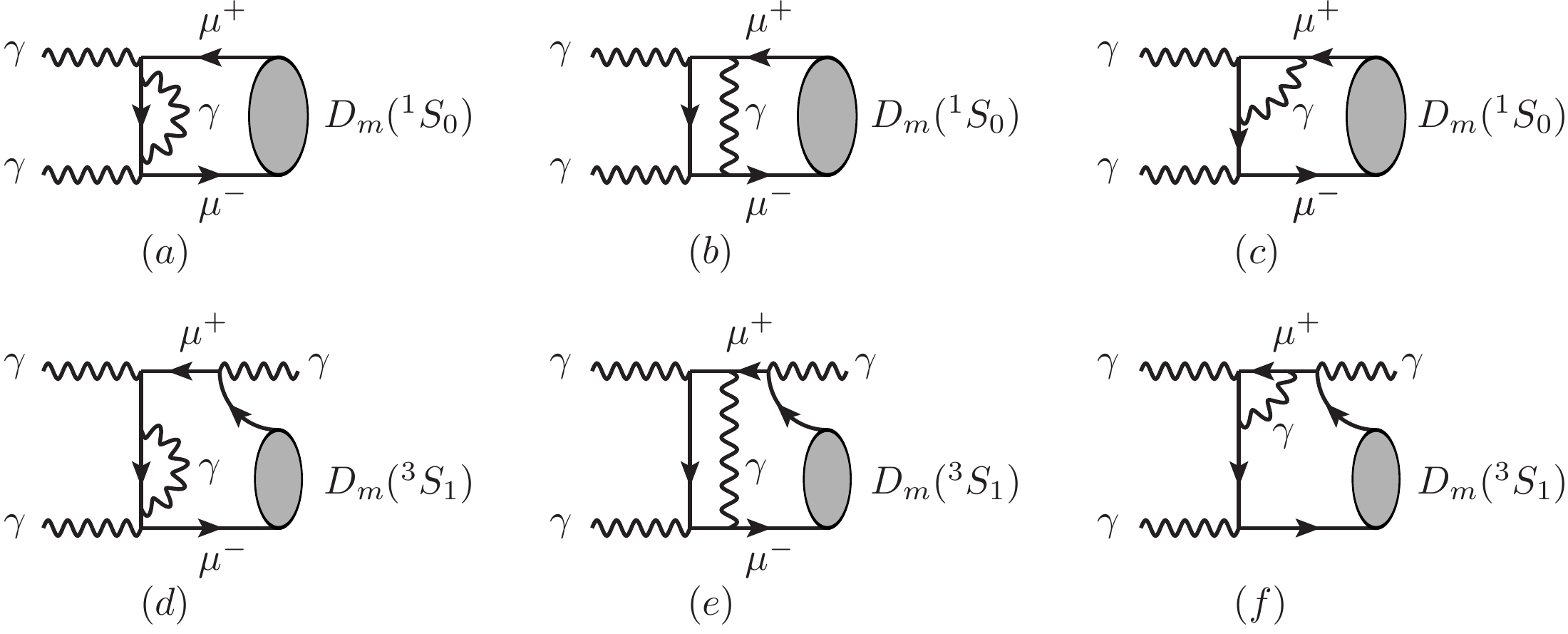}
    \caption{Typical NLO Feynman diagrams of para (the first line) and ortho (the second line) dimuonium photoproduction in photon-photon collision.}\label{fig_nlo}
\end{figure}

In FIG.~\ref{fig_nlo} we show some typical NLO diagrams of $D_m(^1S_0)$ and $D_m(^3S_1)$ production. Due to the restriction of charge conjugation symmetry in QED, the real corrections to both $D_m(^1S_0)$ and $D_m(^3S_1)$ production vanish. In practice, there are 12 diagrams for $D_m(^1S_0)$ and 84 diagrams for $D_m(^3S_1)$. Note, considering of symmetries within the Feynman diagrams, in the end the real number of the contributing and topologically independent diagrams are greatly reduced. In the calculation, following software packages are employed:
\begin{itemize}
    \item FeynArts \cite{Hahn:2000kx}: for generating Feynman diagrams
    \item FeynCalc \cite{Shtabovenko:2020gxv}: for generating analytical amplitude expressions
    \item FIRE6 \cite{Smirnov:2019qkx}: for loop integral reduction
    \item CUBA \cite{Hahn:2004fe}: for numerical phase space integral  
\end{itemize}

By adopting dimensional regularization (DR), divergences in loop diagrams can be further split into
\begin{align}
|\mathcal{M}_{\text{Loop}}|^2 = \mathcal{C}_{UV}\frac{1}{\varepsilon_{UV}} + \mathcal{C}_{IR}\frac{1}{\varepsilon_{IR}} + \mathcal{C}_{fin} + \frac{2\alpha m}{\lambda}|\mathcal{M}_{\text{LO}}|^2 \ .
\end{align}
Here, the first three terms in the right hand side signify UV divergent, IR divergent, and finite terms, respectively. The last term contains the Coulomb divergence, with $\lambda$ being introduced as a spurious photon mass to regulate it. The UV divergence can be readily renormalized with renormalization constants:
\begin{align}
&\delta Z_2^{\rm OS}=-\frac{\alpha}{4\pi}\left[\frac{1}{\varepsilon_{\rm {UV}}}+\frac{2}{\varepsilon_{\rm {IR}}}-3\gamma_E+3\ln\frac{4\pi\mu_r^2}{m^2}+4\right]\ ,\nonumber\\
&\delta Z_m^{\rm OS}=-3\frac{\alpha}{4\pi}\left[\frac{1}{\varepsilon_{\rm UV}}-\gamma_E+\ln\frac{4\pi\mu_r^2}{m^2} +\frac{4}{3}\right]\ , \nonumber\\
&\delta Z_3^{\rm OS}=0\ ,\nonumber\\
&\delta Z_e^{\rm OS}=-\frac{1}{2}\delta Z_3=0\ .
\label{cts}
\end{align}
The IR divergences appearing in different loops may counteract exactly with each other. Coulomb divergences are cancelled with Coulomb terms in the corrections of wave function in NRQED formalism. 

In the end of the calculation, we have the NLO analytic expressions for $D_m(^1S_0)$ and $D_m(^3S_1)$ production. For $^1S_0$ state, the expression writes concisely:
\begin{align}
    |\mathcal{M}_{\text{NLO}}(\gamma+\gamma\to D_m(^1S_0))|^2 = \left(1+\frac{\alpha}{\pi}\frac{\pi^2-20}{4}\right)|\mathcal{M}_{\text{LO}}(\gamma+\gamma\to D_m(^1S_ 0))|^2\ ,
\end{align}
while for $^3S_1$ state the expression is a bit lengthy, which is presented in the Appendix for cross check and phenomenological use. Note, the NLO correction for $^1S_0$ state is obviously negative, though tiny. In fact, the NLO correction for $^3S_1$ state is negative as well, which is in compatible with the NLO calculation of $^3S_1$ decay to three photons \cite{dEnterria:2022alo}.

\section{Numerical Results}

The total cross section of the $^1S_0$ state can be expressed as:
\begin{align}
    \sigma(^1S_0) = \int dx_1dx_2 f_{\gamma}(x_1)f_{\gamma}(x_2) &\times \frac{1}{2 x_1x_2 S}\left(\frac{1}{2\times2}\sum_{r_1,r_2\in (+,-)}|\mathcal{M}|^2\right) |\Psi(0)|^2\nonumber\\
    &\times \frac{\pi}{M}\delta(\sqrt{x_1x_2 S}-M)\ 
\end{align}
with $r_1$ and $r_2$ being the polarizations of initial photons. The $\delta$-function in the second line comes from the single final state phase space integration, and it will be integrated out by one of the EPA integrations, which gives:
\begin{equation}
    \sigma_{\text{LO}}(^1S_0) = 2\pi^2\alpha^5\int dx_1 f_{\gamma}(x_1)f_{\gamma}\left(\frac{4m^2}{x_1 S}\right)\frac{\theta(x_1 S -4m^2)}{x_1 S}
\end{equation}
and
\begin{equation}\label{1S0_NLO_expression}
\sigma_{\text{NLO}}(^1S_0) = \left(1+\frac{\alpha}{\pi}\frac{\pi^2-20}{4}\right)\sigma_{\text{LO}}(^1S_0)\approx 0.9941\, \sigma_{\text{LO}}(^1S_0)\ .
\end{equation}
Here, $\theta(x)$ is the step function. 

\begin{figure}[htb]
    \centering
    \subfigure[]{
    \includegraphics[width=0.45\textwidth]{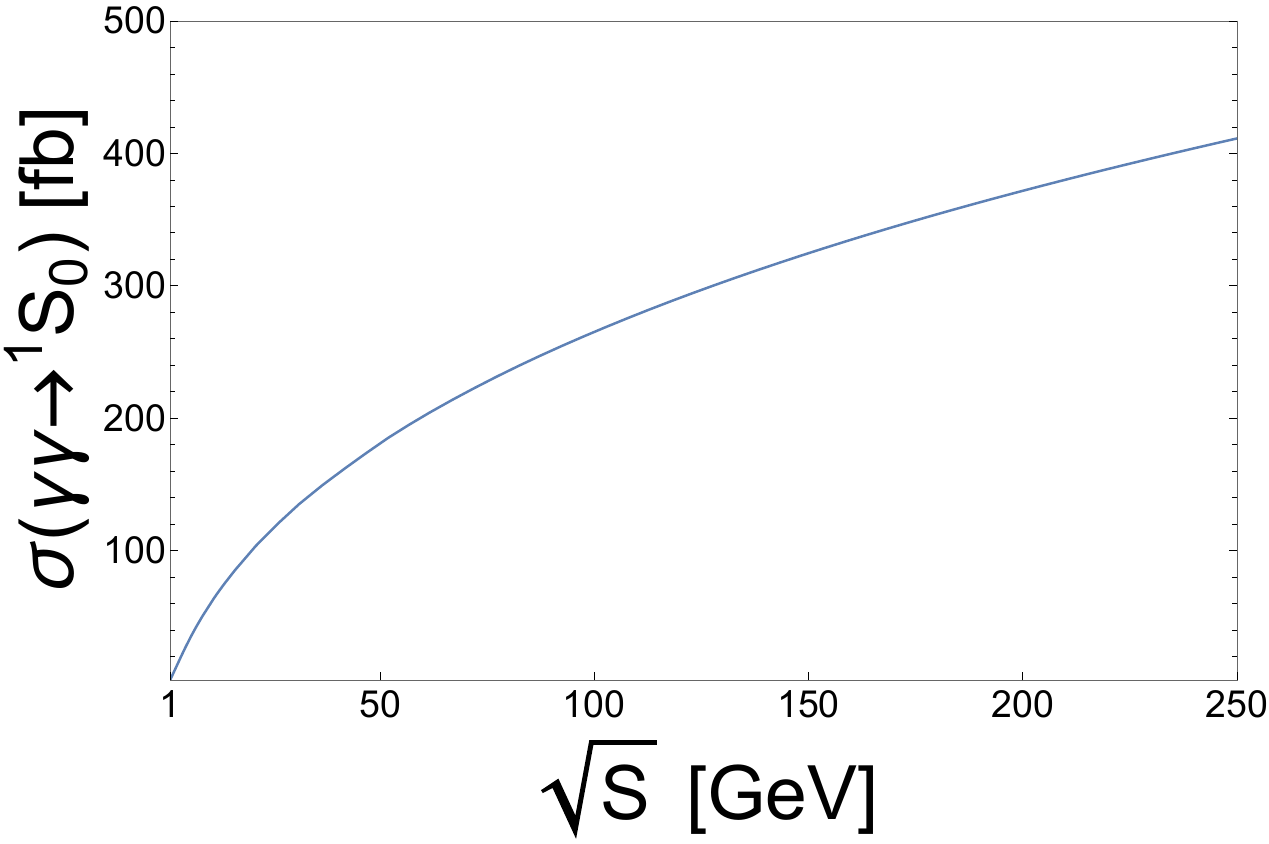}} \hspace{1 cm}
    \subfigure[]{
    \includegraphics[width=0.45\textwidth]{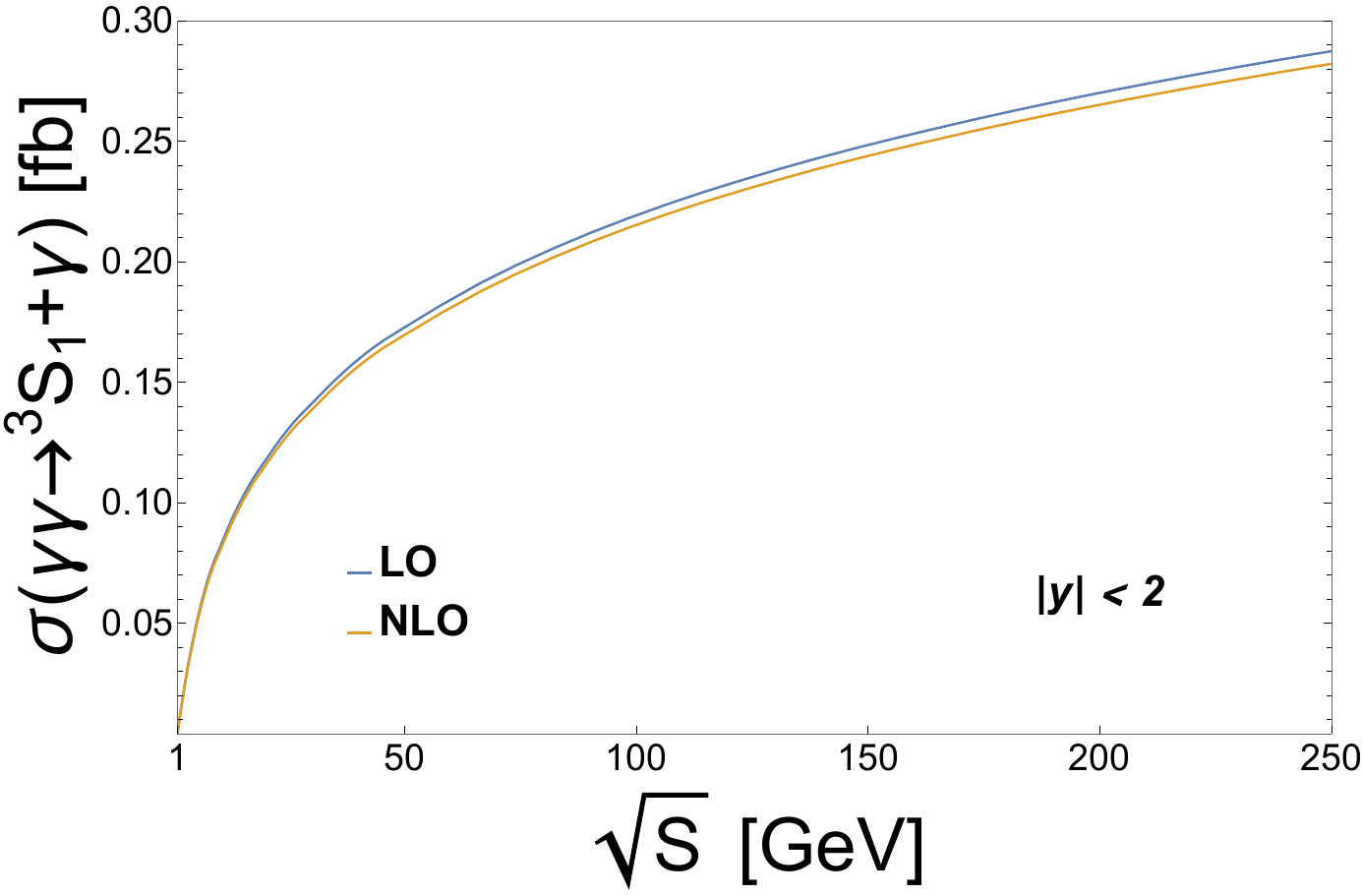}}
    \caption{The NLO total cross sections as a function of the $e^+e^-$ center-of-mass energy $\sqrt{S}$.\@ (a) for $D_m(^1S_0)$ production process and (b) for $D_m(^3S_1)$ production process. }\label{dists}
\end{figure}

In our numerical calculation, the parameter inputs are taken from Particle Data Group (PDG)~\cite{ParticleDataGroup:2024cfk}, which gives muon mass as $m_\mu=105.658\ \rm{MeV}$ and the fine-structure constant as $\alpha = 137.036^{-1}$.
FIG.~\ref{dists} presents the total cross sections for $\gamma\gamma\to D_m(^1S_0)$ and $\gamma\gamma\to D_m(^3S_1)+\gamma$ processes as functions of the $e^+e^-$ collide center-of-mass (c.m.) energy $\sqrt{s}$, computed at NLO order with the EPA. 
From Eq.~\ref{1S0_NLO_expression}, it is noticeable that the NLO correction to $^1S_0$ state is tiny, contributing only about $0.59\%$ of the LO result, i.e. $-0.59\%\, \sigma_{\text{LO}}(^1S_0)$. Hence, in FIG.~\ref{dists} (a) only the NLO result is presented. For the $^3S_1$  dimuonium state, the NLO correction ranges from approximately $(1.56\%\sim 1.82\%)\, \sigma_{\text{LO}}(^3S_1)$, increase with the c.m. energy, and both LO and NLO results are shown in FIG.~\ref{dists} (b).  

In the calculation, for $D_m(^1S_0)$ production, since $e^+e^-$ c.m. frame is adopted, we integrated over the full phase space. For  $D_m(^3S_1)$ production, based on the typical detector acceptance, a rapidity cut $|y|<2$ on the $^3S_1$ state is applied. 
Notice, both channels exhibit a smooth increase from dimuon threshold followed by an energy-independent plateau in considered energy range, reflecting the dominance of the quasi-real photon flux over the subleading perturbative corrections. The $D_m(^3S_1)$ production rate is much smaller than that of $D_m(^1S_0)$ due to the extra photon emission, but their $\sqrt{S}$ dependences are similar.
Consequently, the cross sections and expected event rates at currently running and forthcoming colliders are listed in Table~\ref{tab}. Note, in evaluating the cross sections and event rates, only the dominant decay channels are considered. Following Ref.~\cite{Karshenboim:1998am}, para dimuonium primarily decays via $D_m(^1S_0)\to\gamma\gamma$ channel, with a branching fraction $\mathcal{B}(D_m(^1S_0)\to \gamma\gamma)\approx 98.5\%$, whereas ortho dimuonium mainly decays via $D_m(^3S_1)\to e^+ e^-$ process, with a branching rate $\mathcal{B}(D_m(^3S_1)\to e^+ e^-)\approx 86\%$.

\begin{figure}[htb]
    \includegraphics[width=0.6\textwidth]{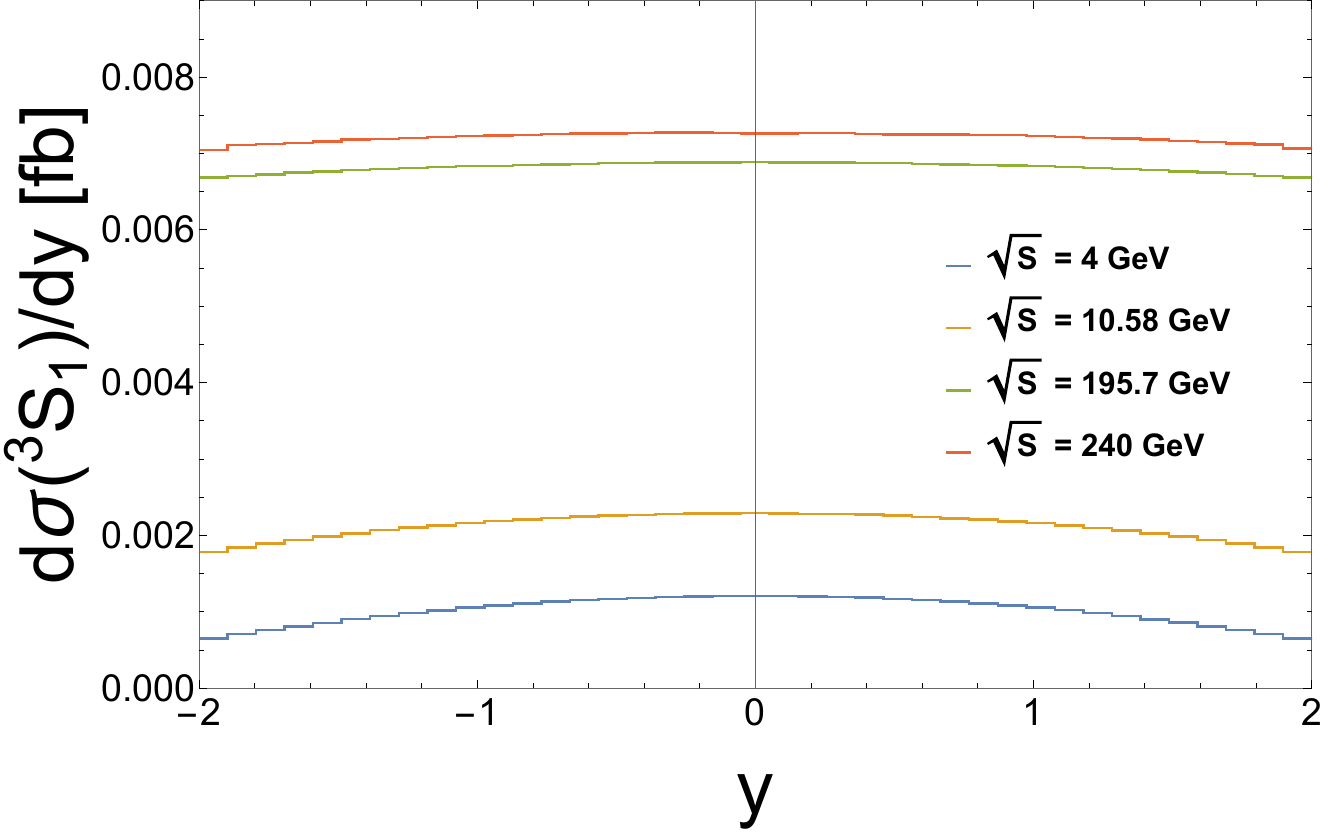}
    \caption{Rapidity distributions for the photoproduction of $^3S_1$ dimuonium in $e^+e^-$ colliders.}\label{ydist}
\end{figure}

The differential cross section distribution of $D_m(^3S_1)$ production versus rapidity $y$ at various $\sqrt{S}$ are presented in FIG.~\ref{ydist}. We evaluate four representative $e^+e^-$ center-of-mass energies for colliders listed in Table \ref{tab}.
All curves exhibit a symmetry around $y = 0$ and vary mildly over the full range, remaining nearly flat with slight central deviations. With the increase of $\sqrt{S}$, the overall normalization rises significantly while the overall shape changes moderately.

\begin{table}[htb]
\centering
\caption{NLO cross sections (in fb) and event rates per year (except for LEP II which shows the event rate across 1996-2000 with DELPHI measurements~\cite{Abdallah:2006yg}) of $D_m(^1S_0)$ and $D_m(^3S_1)$ states photoproduction in $e^+e^-$ collision. }
\label{tab}
\begin{tabular}{lcc|cc|cc}
\toprule[2pt]
\multirow{2}{*}{\textbf{Collider}} & 
\multirow{2}{*}{\begin{tabular}{c}\textbf{c.m. Energy}\\$[{\rm GeV}]$\end{tabular}} &
\multirow{2}{*}{\begin{tabular}{c}\textbf{Luminosity}\end{tabular}} &
\multicolumn{2}{c|}{~\textbf{Cross section}~} &
\multicolumn{2}{c}{\textbf{Event}} \\
\cmidrule(lr){4-5} \cmidrule(lr){6-7}
 & & & $D_m(^1S_0)$ & $D_m(^3S_1)$ & $D_m(^1S_0)$ & $D_m(^3S_1)$\\
\midrule
\toprule[1pt]
BEPC II~\cite{ParticleDataGroup:2024cfk}        
 & 3.78   & $1\times10^{33}\ {\rm cm^{-2}s^{-1}}$ & 21.1 & 0.0318 & 666 & 1 \\
STCF~\cite{Achasov:2023gey}           
 & 4.0    & $5\times10^{34}\ {\rm cm^{-2}s^{-1}}$ & 22.6 & 0.0336 & 35661 & 53 \\
SuperKEKB~\cite{ParticleDataGroup:2024cfk}      
 & 10.58  & $4.71\times10^{34}\ {\rm cm^{-2}s^{-1}}$ & 59.9 & 0.0704 & 88915 & 104 \\
LEP II~\cite{Abdallah:2006yg}        
 & 195.7  & $617\ {\rm pb^{-1}}$ & 362. & 0.227 & 223 & - \\
CEPC/FCC-ee~\cite{ParticleDataGroup:2024cfk}    
 & 240    & $5\times10^{34}\ {\rm cm^{-2}s^{-1}}$ & 398. & 0.240 & 627136 & 378 \\
\bottomrule[1pt]
\end{tabular}
\end{table}

\section{Conclusion and Remarks}

In this work, we calculate the NLO QED corrections to para and ortho dimuonium production in photon-photon collision.
Based on the NLO calculation results, one can keep on holding the confidence of Gargiulo, Meco and Palmisano's \cite{Gargiulo2025} prediction on the feasibility of observing $D_m(^1S_0)$ in Belle II experiment. The NLO QED corrections to both spin-singlet and triplet states are negative. To singlet, the correction is a phenomenologically negligible constant, while to triplet, the correction is in proportion to
the colliding energy, that is from $1.56\%$ to $1.82\%$ as energy goes from BEPC II to CEPC. From our calculation, the most possible finding machine for dimuonium at present is Belle II in electron-positron collision, as Ref. \cite{Gargiulo2025} analyzed, while in the future STCF and CEPC/FCC may have a great expectation in observing them.

Finally, the dimuonium is a pure QED atom-like bound system. From theoretical point of view, it should and could be produced and observed in present high energy experiment, for the investigation of QED delicacy. Calculation implies, though photon-photon interaction is a very possible mechanism for dimuonium production and detection, that to discover other attainable means are still necessary for this aim.

\vspace{1.0cm} {\bf Acknowledgments}

Authors are appreciated to Ms. Yuan-Hui Zhu for organization help of this research group. This work was supported in part by the National Natural Science Foundation of China(NSFC) under the Grants 12475087, 12235008 and 12235001, and by University of Chinese Academy of Sciences.
\newpage
\appendix{\textbf{Appendix}}

Following is the squared matrix element of $\gamma+\gamma\to {Q_m}[\mu^+ \mu^-] + X$ process at the NLO QED. To make the expression concise, the Mandelstam variables $s$ and $t$ are dimensionless normalized, i.e. $s \rightarrow s/M_{Q_m}$ and $t \rightarrow t/M_{Q_m}$. In practice, $s>1,\ t<0,\ u=1-s-t<0$, and the real part of analytical expression is simplified based on these conditions. Since the variables $s,\ t,\ u$ are symmetric in $|\mathcal{M}|^2$, by replacing $s,\ t$ in $f(s,t)$ and sum, we have  
\begin{align}
\frac{\mathcal{M}^{*}_{\rm tree}\mathcal{M}_{\rm oneloop}}{e^8}&=c_0+f(s,t) + f(t,s)\nonumber\\
&+f(s,1-s-t)+f(t,1-s-t)+f(1-s-t,s)+f(1-s-t,t)\ .
\end{align} 
Since there are no log or dilog terms in $c_0$, we present it in the $s,\ t,\ u$ symmetric form for simplicity, rather in $f(s,t)$. The analytical expression of $f(s,t)$ writes:
\begin{align}
f(s,t)=f_{\log}+\displaystyle\sum_i c_if_i\ ,
\end{align}
where 
\begin{align}
f_1&=\log(2-2s),\ f_2=\log[1 - 2s + 2\sqrt{s(s-1)}]\ ,\nonumber\\
f_3&=f_2^2,\ f_4=\dlog(2s-1),\ f_5=\dlog(2t-1)\ ,\nonumber\\
f_6&=\dlog(1-1/s)+\dlog(1-1/t)\ ,\nonumber\\
f_7&=\frac{1}{k}\left\{\dlog\left(-2\frac{k+st}{1-s-t}\right)-\dlog\left(2\frac{k-st}{1-s-t}\right)\right.\nonumber\\
&+\dlog\left(2+2\frac{k+st}{1-s-t}\right)-\dlog\left(2-2\frac{k-st}{1-s-t}\right)\nonumber\\
&+\dlog\left[1-\frac{1}{1-2s+2\sqrt{s(s-1)}}\left(1-2\frac{k-st}{1-s-t}\right)\right]\nonumber\\
&-\dlog\left[1-\frac{1}{1-2s+2\sqrt{s(s-1)}}\left(1+2\frac{k+st}{1-s-t}\right)\right]\nonumber\\
&+\dlog\left[1-\frac{1}{1-2s-2\sqrt{s(s-1)}}\left(1-2\frac{k-st}{1-s-t}\right)\right]\nonumber\\
&\left.-\dlog\left[1-\frac{1}{1-2s-2\sqrt{s(s-1)}}\left(1+2\frac{k+st}{1-s-t}\right)\right]\right\} \ , \nonumber\\
f_8&=\frac{2}{k}\left\{2\dlog\left(2s-\frac{2k}{1-t}\right)-2\dlog\left(2s+\frac{2k}{1-t}\right)\right.\nonumber\\
&+\dlog\left[1-\frac{1-t}{1-s}\left(1-2s-\frac{2k}{1-t}\right)\right]-\dlog\left[1-\frac{1-t}{1-s}\left(1-2s+\frac{2k}{1-t}\right)\right]\nonumber\\
&+\dlog\left[1-\frac{1}{1-2s+2\sqrt{s(s-1)}}\left(1-2s-\frac{2k}{1-t}\right)\right]\nonumber\\
&-\dlog\left[1-\frac{1}{1-2s+2\sqrt{s(s-1)}}\left(1-2s+\frac{2k}{1-t}\right)\right]\nonumber\\
&+\dlog\left[1-\frac{1}{1-2s-2\sqrt{s(s-1)}}\left(1-2s-\frac{2k}{1-t}\right)\right]\nonumber\\
&\left.-\dlog\left[1-\frac{1}{1-2s-2\sqrt{s(s-1)}}\left(1-2s+\frac{2k}{1-t}\right)\right]\right\}\ . 
\end{align}
Here, $k=-\sqrt{s t (1-s) (1-t)}$, $f_7(s,t)+f_7(t,s)$ comes from the four-point Passarino-Veltman function $D_0[1,0,0,0,s,t,\frac{1}{4},\frac{1}{4},\frac{1}{4},\frac{1}{4}]$,  and $f_8$ from $D_0[\frac{1}{4},0,0,\frac{s}{2}-\frac{1}{4},\frac{t}{2}-\frac{1}{4},s,0,\frac{1}{4},\frac{1}{4},\frac{1}{4}]$.

The coefficients $c_i$ and $f_{\log}$ are
\begin{align}
c_0&=\frac{-16}{u^4 (1 - u + v)^4}[30 u^8 - 16 v (1 + v)^3 + 2 u v (1 + v)^2 (28 + 5 v)\nonumber\\
& - 6 u^7 (20 + 19 v) + u^6 (255 + 336 v + 100 v^2) +  u^4 (271 + 603 v + 31 v^2 - 72 v^3)\nonumber\\
& - u^5 (345 + 568 v + 58 v^2 + 32 v^3) + u^2 (16 - 29 v - 54 v^2 - 5 v^3 + 4 v^4) \nonumber\\
&- u^3 (107 + 268 v + 77 v^2 - 6 v^3 - 32 v^4)]
+\frac{8}{\pi^2 u^3 (2 u-1) (2 u - 4 v-1)(1 - u + v)^3}
[1504 u^8\nonumber\\
& - 32 u^7 (188 + 151 v) + 4 u^6 (2339 + 3328 v + 1008 v^2) \nonumber\\
&+ 2 u v (147 + 425 v - 216 v^2 - 464 v^3) +  4 v (1 + v)(1-16v^2) \nonumber\\
&- 4 u^5 (1753 + 2504 v + 408 v^2 + 192 v^3) + u^4 (2489 - 2500 v - 13684 v^2 - 4992 v^3)\nonumber\\
& - 2 u^3 (155 - 3140 v - 8846 v^2 - 3652 v^3 - 384 v^4) \nonumber\\&- u^2 (11 + 2542 v + 7273 v^2 + 1380 v^3 - 1728 v^4)]\  \\
&\text{with $u=s+t$, $v=st$,}\nonumber\\
c_1 & =\frac{-16}{\pi^2 (1 - 2 s)^2 (s-1)^3 (s - t) (t-1)^2 (s + t)^2(2 s + t-1)}\nonumber\\
& [8 s^9 + s^8 (24 - 88 t)+ 3 t^2 (2t-1)^2(t-1)^2 - 4 s^7 (43 - 111 t + 38 t^2) \nonumber\\
& + s^6 (315 - 827 t + 618 t^2 - 144 t^3) - 4 s^5 (68 - 189 t + 244 t^2 - 127 t^3 + 28 t^4) \nonumber\\
&+  s t (3 - 46 t + 168 t^2 - 265 t^3 + 192 t^4 - 52 t^5) 
 + s^4 (121 - 375 t + 886 t^2 - 906 t^3 + 422 t^4 - 48 t^5)\nonumber\\& 
- s^3 (26 - 109 t + 542 t^2 - 964 t^3 + 761 t^4 - 240 t^5 + 16 t^6) \nonumber\\
&+ s^2 (2 - 22 t + 217 t^2 - 572 t^3 + 661 t^4 - 348 t^5 + 64 t^6)]\ ,\\
c_2 & =\frac{16\sqrt{(s-1) s}}{\pi^2 (s-1)^4 (t-1)^2 (s + t)^2} [s^5 (t-3) + s^4 (5 - 3 t - t^2) + s^3 (3 t - 4 t^3)
 - 2 t (1 - 7 t + 12 t^2 - 6 t^3)\nonumber\\ 
&+ s (1 + 12 t - 36 t^2 + 28 t^3 - 2 t^4) - s^2 (3 + 11 t - 15 t^2 + 2 t^4)]\ ,\\
c_3 & =\frac{8 [3 s^5 + 2 s^4 t + 3 s^2 (4 + 5 t - 2 t^2) - 2 s^3 (6 + 4 t - t^2) - s (3 + 10 t - 9 t^2) - t (t - 1) ]}{\pi^2 (s-1)^4 (t-1) (s + t)}\ , \\    
c_4& =\frac{-16}{\pi^2 (s-1)^4 (s - t) (t-1)^2 (s + t)^2}[2 s^7 + s^6 (2 - 12 t) + s^5 (2t-1)(2t+17) \nonumber\\&+ s^4 (25 - 24 t - 26 t^2 + 12 t^3) - s^3 (19 - 2 t - 51 t^2 + 36 t^3 + 5 t^4)\nonumber\\& + t (2 - 5 t + 4 t^2 + 4 t^3 - t^4) + s^2 (9 + 6 t - 40 t^2 + 27 t^3 + 18 t^4 - t^5) \nonumber\\&-s (2 + 6 t - 20 t^2 + 11 t^3 + 21 t^4 - 6 t^5)]\ , \\    
c_5& =\frac{-8 [s - 3 s^2 - 2 (s - t)^4 + t - s t - 3 t^2 + 2 s t (s + t)]}{\pi^2 (s-1)^2 (s - t) (t-1)^2 (s + t)}\ ,\\    
c_6&=\frac{16 (s^2 + t^2)}{\pi^2 (s-1) (t-1) (s + t)}\ , \\ 
c_7& =\frac{-8 [s^2 (t-2) + 2 s^3 t - 2 (t-1) t + s (2 - 2 t + t^2 + 2 t^3)]}{\pi^2 (s-1) (t-1) (s + t)}\ , \\
c_8& =\frac{-4 s [s^3 (4 t-2) + s^2 (4t-3)(2t-3) - s (9 - 9 t - 7 t^2 + 4 t^3) + 2 (1 + t - 4 t^2 + 2 t^4)]}{\pi^2 (s-1)^2 (t-1)^2 (s + t)}\ ,\\
\text{and}\nonumber\\
f_{\log}& =\frac{-8}{\pi^2 (1 - s) (t-1) (s + t)}[(1 + 3 s^2 + 2 (s+t) (t-1)] \log^2 s\nonumber\\
&+ 2(s - 2 s^2 + t + s t - 2 t^2) \log ~t-2 \log~s~ \log~t(s^2 + t^2)]\nonumber\\
&-\frac{16s^3 \log~s}{\pi^2 (1 - s)^3 (t-1) (s + t)}(4 s +7 t-13)\nonumber\\
& -\frac{16 \log~s}{\pi^2 (1 - s)^3(s + t)}[s^2 (4t-15)- s (8t-7)+2t-1]\ .
\end{align}

\end{document}